# Analysis and Comparison of Time Series of Power Consumption of Sistan and Tehran distribution networks

Masoud Safarishaal,

Technical and Engineering Faculty, Imam Khomeini International University, Qazvin, Iran[1]

masoud.safarishaal66@gmail.com,

**Abstract.** Data presented in the form of time series as its analysis and applications recently have become increasingly important in different areas and domains. Prediction and classification of time-series data play a vital role in multiple fields. In this paper, the time series analysis related to power consumption at 12 o'clock every day in the period of 2012 to 2014 has been compared for two distribution networks of Sistan and one of the four networks of Tehran. By analyzing the power consumption of these two networks, a comparison can be made between these two regions in terms of development and climate difference and the impact of social, industrial and environmental phenomena. The reason for choosing these two networks was to compare a deprived area with an area in the capital. CRP tool software and toolkits have been used to analyze and compare time series, and various tools have been used to compare two time series.

**Keyword:** Time series, Load Forecasting, Power Distribution, and CRP tool software

## 1- Introduction

The analysis of experimental data that have been observed at different points in time leads to new and unique problems in statistical modeling and inference. Data presented in the form of time series as its analysis and applications recently have become increasingly important in different areas and domains. [1-4] Time series analysis pursues two goals, understanding and modeling the stochastic mechanism that leads to the occurrence of a series of observations and predicting the future values of the series based on its past [5]. Load forecasting is a process of predicting the future load demands. It is important for power system planners and demand controllers in ensuring that there would be enough generation to cope with the increasing demand. Accurate model for load forecasting can lead to a better budget planning, maintenance scheduling and fuel management. In this article, two time series related to power consumption at 12 o'clock every day in the period of 2012 to 2014 for the two distribution networks of Sistan and one of the four networks of Tehran have been compared and analyzed with each other. Figure 1 shows the two time series mentioned. In the beginning, the data are normalized. The normality of the data is important because the theory of time series has been developed based on the normality of the data, and if the data is not normal, they should be normalized using different methods. One of the methods of data normalization is the use of logarithmic methods of data [6, 7]. As can be seen from Figure 1, the time series trend related to Tehran is increasing at a higher rate than Sistan, perhaps this difference can be attributed to the faster rate of industrial development, or faster population growth in Tehran. Also, as shown in the figure, the difference in power consumption in the first 6 months of the year with the second half of the year in Sistan is greater, because the weather in Sistan is warmer, and also because there are many more training centers in Tehran, the centers are closed Training in the summer has had a greater impact on reducing the power consumption difference in half a year. Most of Tehran's power consumption is related to industry. The software fits a line on the data to determine the process component. The slope of this line is equal to the trend component. Obviously, if the slope of this line is zero and the line is horizontal, the data has no trend component and is static [8].

In order to better examine the time series, the time series were de-trended with the detrend (data) command. Figure 2 shows the new time series. From now on, we use detrend data for analysis when the goal is to examine two time series together.

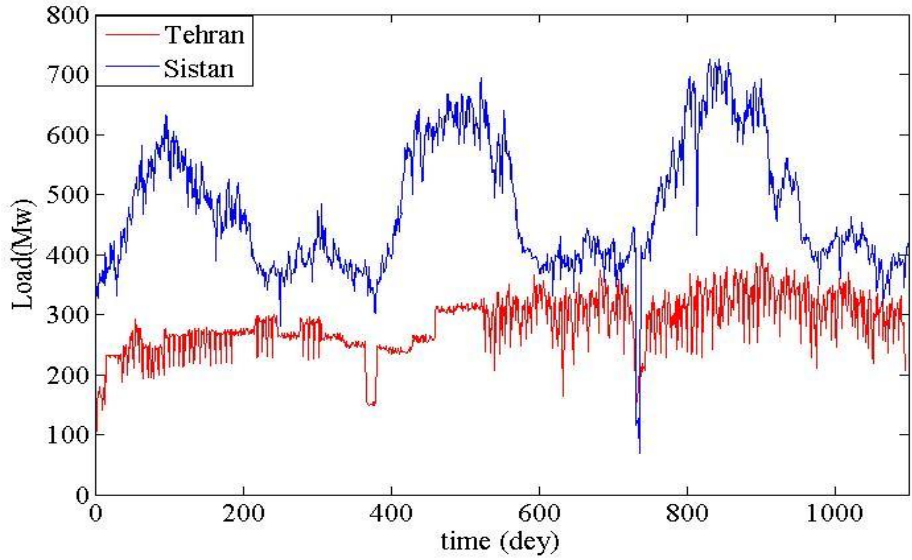

Fig1. Load of Sistan and Tehran

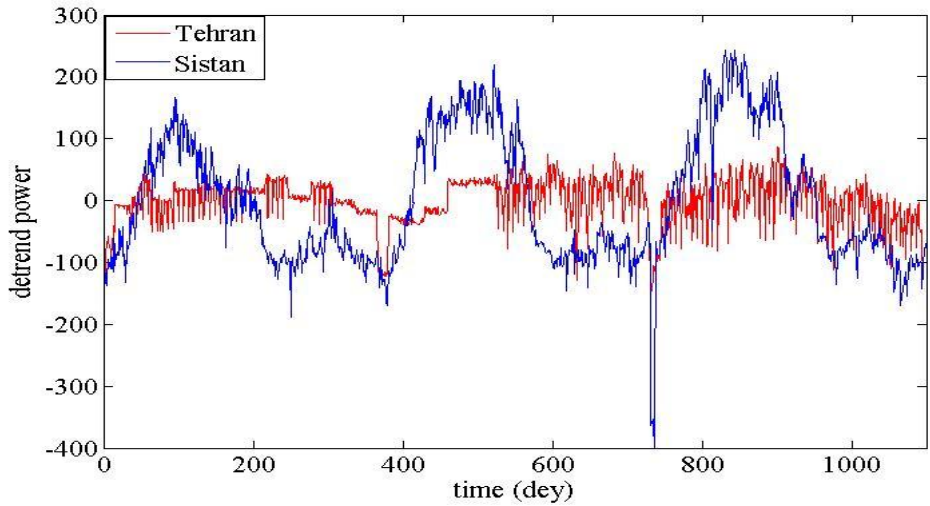

Fig 2. Graph related to the de-trended time series of Sistan power consumption and Tehran power consumption.

## 2- Histogram Diagram

Figure 3 shows the histogram of two networks. The histogram is a graph of classified frequencies that shows the ratio of different states in separate classes (in a bar view). Comparison of the histogram diagram in Figures 3 (a) and (b) shows that the Sistan network power consume is more than Tehran, but a comparison of Figure 3 (c), which shows a detrend comparison of the data shown in a graph, shows that the load fluctuations during the year are very small and this is very convenient in terms of operation. So, investors in the manufacturing sector are more willing to invest in such systems.

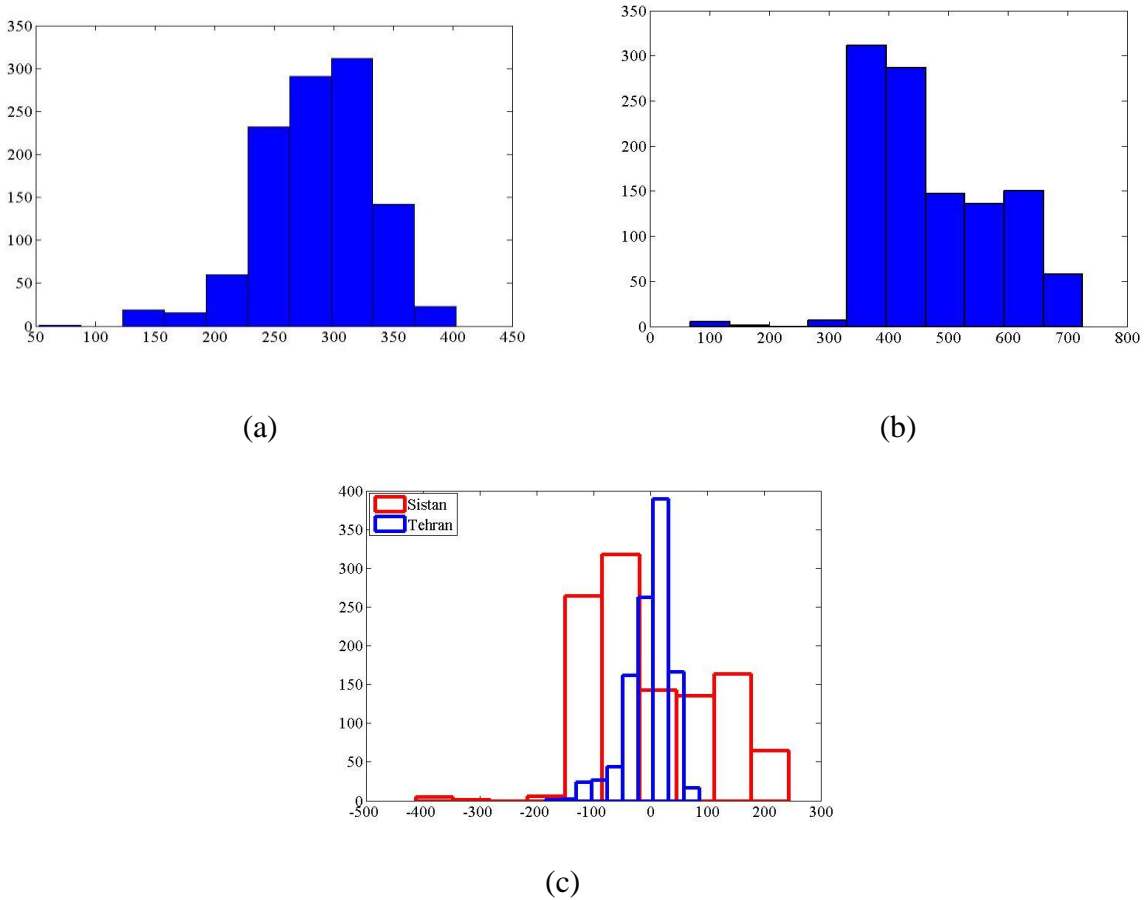

(a)　　　　　　　　　　　　　　　　　　(b)

(c)

Fig. 3 **hist** diagram: a. Tehran b. Sistan c detrend Tehran and Sistan

### 3- False Nearest Neighbor

The False Nearest Neighbor (FNN) graph is shown in Figure 4 for both time series. This chart can be executed by applying the fnn (data) command in MATLAB software. This method is used to determine the time series dimension. Sistan time series has dimension 8 and Tehran time series has dimension 7, which indicates that Sistan has less predictability and it is difficult to predict.

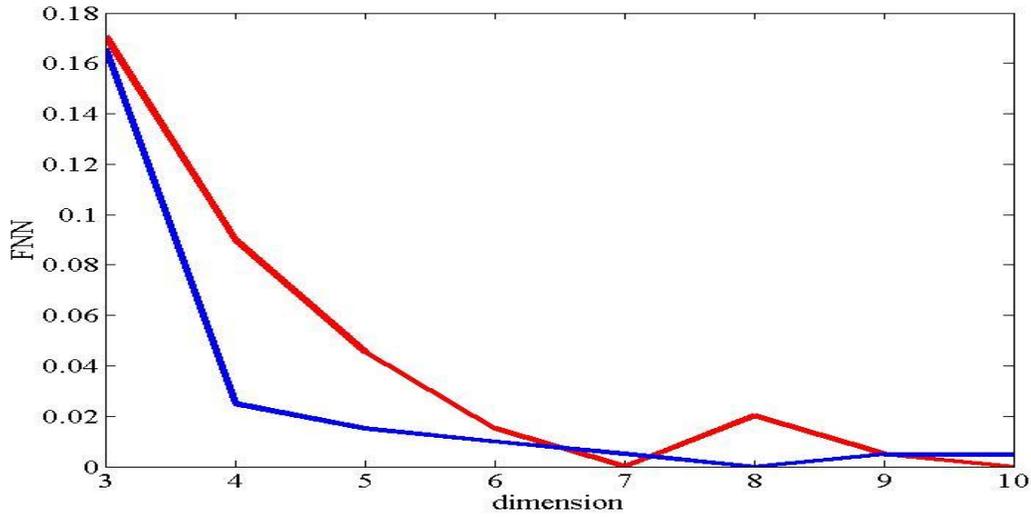

Fig. 4 False Nearest Neighbor (FNN)

### 4- Mutual Information (MI)

Figure 5 is a diagram of the mi (data) command. Which expresses the mutual information method. This method is used to determine the amount of delay. According to the form of delay, it is 6 for Sistan and 5 for Tehran. The first length that the chart is minimized indicates the delay. The phase body diagram for the time series in Figure 7 shows that the time series behavior was chaotic, so they have little predictability.

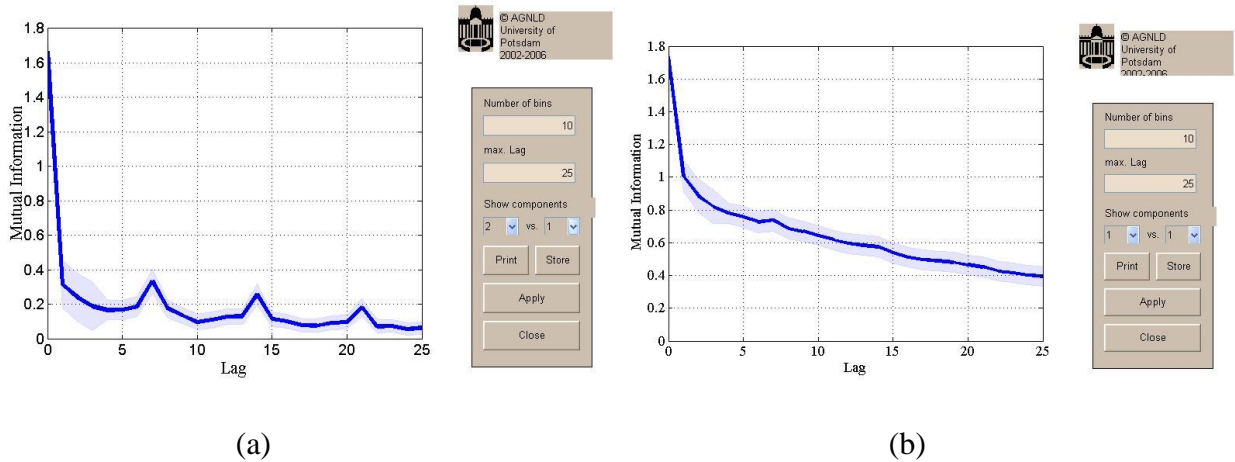

(a)          (b)

Fig 5. MI. (A) Tehran (b) Sistan

## 5- Cross Recurrent Plot

The cross recurrent plot diagram in Figure 6 shows the chaotic nature of both systems, although the number of parallel lines for the time series related to Tehran power consumption is higher, and as indicated in the previous methods, this method is also confirmed Shows that Tehran time series is more predictable than Sistan time series. Despite these large squares in the figure related to the Sistan time series, it shows the seasonal behavior of Sistan power consumption, which was previously guessed according to the time series diagram.

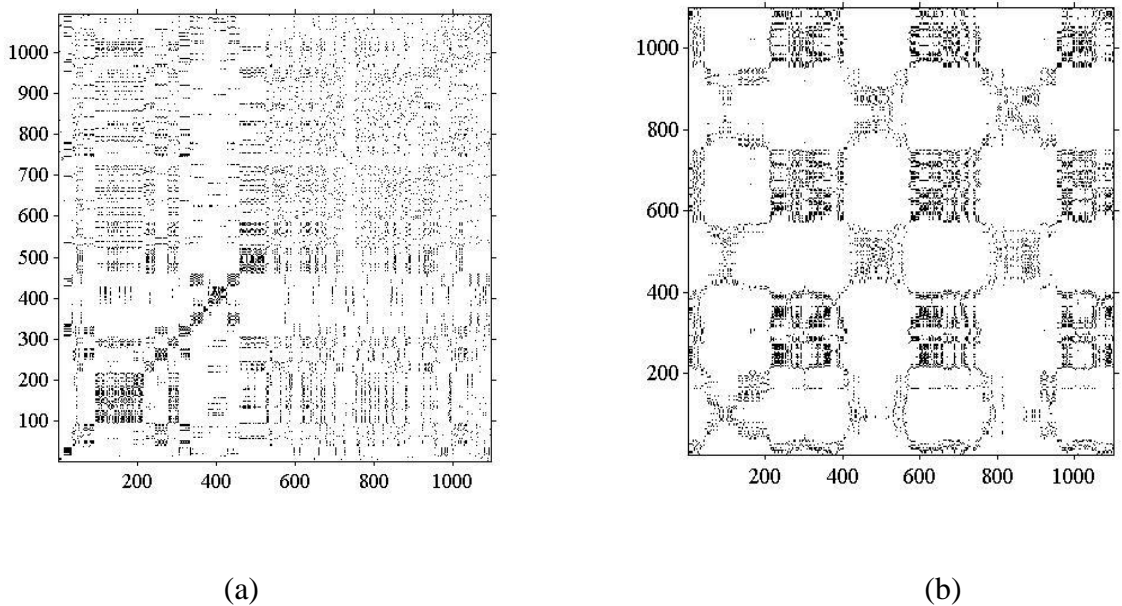

(a) (b)

Fig6. Diagram of cross recurrent plot (a) Tehran (b) Sistan

## 6- xcf diagram

In figure 7. The xcf diagram for the two time series is listed. This command shows the correlation of the data with each other. As can be seen from the figure, the correlation of data in Tehran is very high and this shows the closeness of the data. Naturally, the proximity of the data also makes it easier to predict.

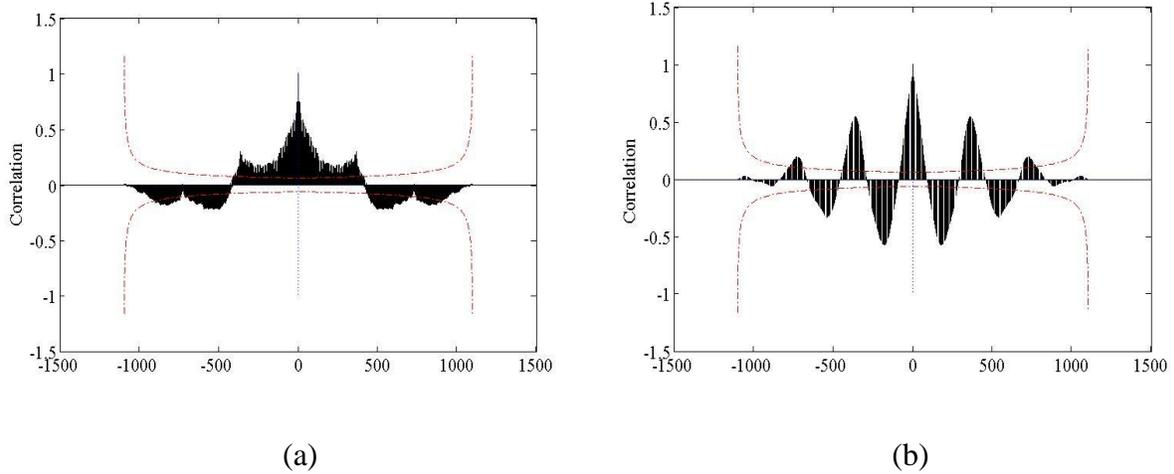

(a)             (b)

Fig. 7 xcf diagram: (a) Tehran (b) Sistan

## 7- Power Spectrum Density

Power spectrum density diagrams with the psd (data) command are shown in Figure 8. The density spectral integral plots the average signal strength over a range of frequencies.

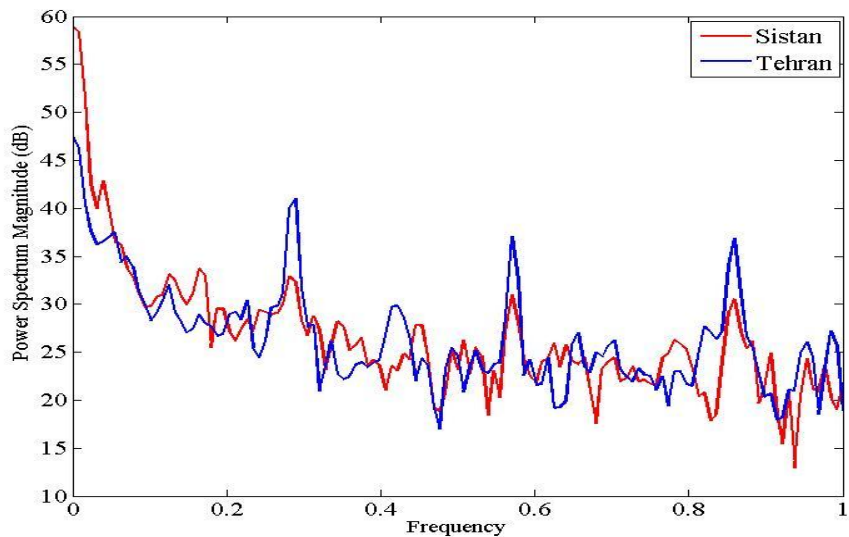

Fig. 8: psd diagram of two time series

In figure 9. Phase body diagram for the two time series mentioned in the 3D display is depicted.

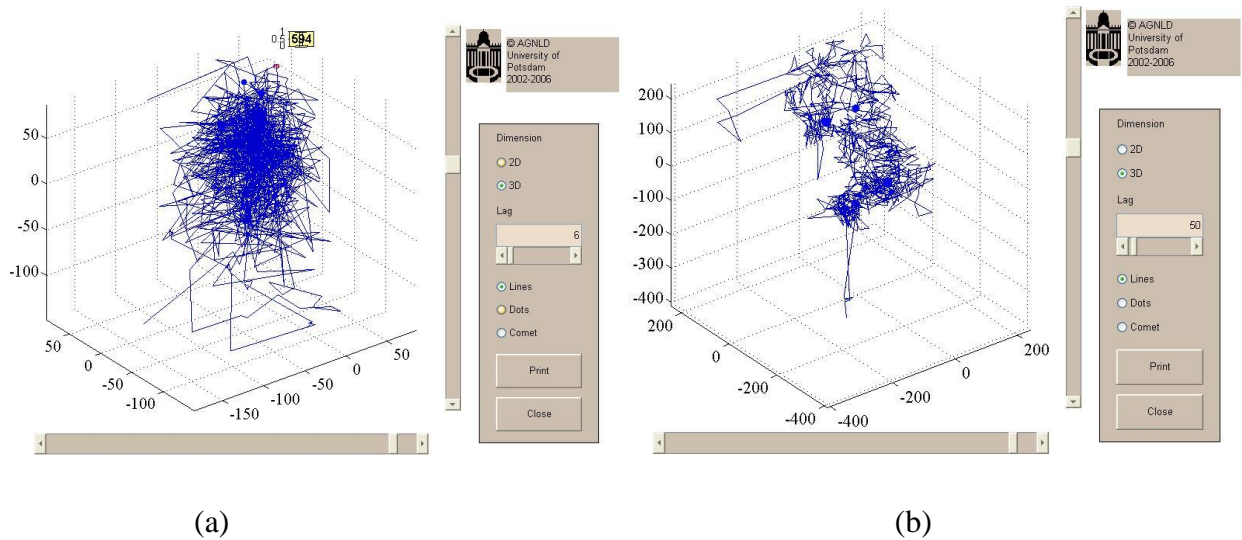

(a)                                                (b)

Fig.9 Phase body diagram for a) Sistan and b) Tehran time series

## 8- Conclusion

In this article, two time series related to power consumption at 12 o'clock every day in 2012 and 2014 for two distribution networks of Sistan and one of the four regions of Tehran were studied and analyzed and compared. The results of seasonal power in Sistan and impact most of it indicated the weather conditions, which is due to the fact that most of the power consumption in Sistan is the power consumption of household appliances, while in Tehran, due to the industrial nature of the city, power consumption is less dependent on water conditions. And it has air and therefore is more predictable. Also, the more upward trend in power consumption in Tehran is due to the faster development and increase in population, while Sistan, as a deprived area, has not had many upward changes during the two years. The difference in power consumption in different seasons in Sistan is quite more noticeable than in Tehran.

## 9- Reference